\documentclass[pdflatex,iicol,sn-mathphys]{sn-jnl}

\usepackage{xspace}
\usepackage{csquotes}

\jyear{2022}

\theoremstyle{thmstyleone}%
\theoremstyle{thmstyletwo}%

\theoremstyle{thmstylethree}%

\newcommand{\pdfs}{\textit{p.d.f.s}\xspace}

\newcommand{\eq}[1]{Eq.\,\ref{eq:#1}\xspace}
\newcommand{\fg}[1]{Fig.\,\ref{fg:#1}\xspace}
\newcommand{\tb}[1]{Tab.\,\ref{tb:#1}\xspace}

\graphicspath{{fig/}}

\begin{document}

\title[Article Title]{A fast and stable approximate maximum-likelihood method for template fits}

\author*[1]{\fnm{Hans} \sur{Dembinski}}\email{hans.dembinski@tu-dortmund.de}

\author[2]{\fnm{Ahmed} \sur{Abdelmotteleb}}\email{ahmed.abdelmotteleb@warwick.ac.uk}

\affil*[1]{\orgdiv{Department of Physics}, \orgname{TU Dortmund}, \orgaddress{\street{August-Schmidt-Straße}, \city{Dortmund}, \postcode{44227}, \country{Germany}}}

\affil[2]{\orgdiv{Department of Physics}, \orgname{University of Warwick}, \orgaddress{ \city{Coventry}, \postcode{CV4 7AL}, \country{United Kingdom}}}

\abstract{Barlow and Beeston presented an exact likelihood for the problem of fitting a composite model consisting of binned templates obtained from Monte-Carlo simulation which are fitted to equally binned data. Solving the exact likelihood is technically challenging, and therefore Conway proposed an approximate likelihood to address these challenges. In this paper, a new approximate likelihood is derived from the exact Barlow-Beeston one. The new approximate likelihood and Conway's likelihood are generalized to problems of fitting weighted data with weighted templates. The performance of estimates obtained with all three likelihoods is studied on two toy examples: a simple one and a challenging one. The performance of the approximate likelihoods is comparable to the exact Barlow-Beeston likelihood, while the performance in fits with weighted templates is better. The approximate likelihoods evaluate faster than the Barlow-Beeston one when the number of bins is large.}




\maketitle

\section{Introduction}\label{sec1}

Barlow and Beeston \cite{Barlow:1993dm} describe the exact likelihood for the problem of fitting a composite model consisting of binned templates obtained from Monte-Carlo simulation which are fitted to equally binned data. In this scenario, the component \pdfs are not available in parameterized form, but are obtained implicitly from simulation. The shapes of the templates are not exactly known since the simulated sample is finite. Therefore, the value of each template in each bin is a nuisance parameter, constrained through Poisson statistics by the simulated sample.

The exact likelihood has many nuisance parameters; one for each component in each bin. Since the number of bins can be large (especially when distributions are multidimensional), Barlow and Beeston describe an algorithm to estimate the nuisance parameters implicitly by solving a non-linear equation per bin for a given set of yields. Since nuisance parameters are implicitly found, only the yields remain as external parameters, which are found numerically in the usual way, for example, with the MIGRAD algorithm in MINUIT \cite{James:1975dr}.

This approach is elegant, but also has drawbacks. It was observed by Conway \cite{Conway:2011in} that the finite accuracy of the non-linear solver may introduce discontinuities in the likelihood that confuses MIGRAD and similar algorithms. This is rarely an issue in practice, but it can lead to fits that fail to converge or fits that produce incorrect uncertainty estimates. Furthermore, solving a non-linear equation per bin adds a non-negligible computational cost.

Conway proposed a simplified treatment \cite{Conway:2011in} to address these issues. The exact likelihood is replaced by a simplified one where the uncertainty in the template is captured by a multiplicative factor, which in turn is constrained by a Gaussian penalty term. This introduces only one nuisance parameter per bin instead of one per component per bin, and it allows one to estimate each nuisance parameter by solving a simple quadratic equation. The computation of the simplified likelihood does not suffer from numerical instabilities and is faster.

Conway did not derive the simplified likelihood rigorously from the exact likelihood of Barlow and Beeston. Such a derivation is therefore attempted here, motivated by the wish to gain insight in which limit Conway's likelihood is an adequate proxy of the exact likelihood. This exercise lead to the discovery of another approximate likelihood that offers the same technical benefits, whilst treating simulation and data symmetrically. Because of these properties, the new approximate likelihood is expected to perform better than Conway's in cases where the simulated sample is small.

All likelihoods derived in this paper are transformed so that the minimum value is asymptotically chi-square distributed, following the approach of Baker and Cousins \cite{Baker:1983tu}. This allows one to use the minimum value as a goodness-of-fit test statistic and has a stabilizing effect on gradient-based minimization algorithms like MIGRAD. Furthermore, the approximate likelihoods are generalized to handle weighted templates, based on results from Bohm and Zech \cite{Bohm:2013gla}. In the last part of the paper, fits to toy examples with the approximate likelihoods are compared to the fits with the fits using the Barlow-Beeston likelihood.

\section{Transformed likelihood as a goodness-of-fit test statistic}\label{sec2}

Baker and Cousins \cite{Baker:1983tu} note that likelihoods for binned data can be transformed such that the minimum value doubles as an asymptotically chi-square-distributed test statistic, $Q$. The following monotonic transformation is applied to the likelihood $\mathcal L$, without loss of generality given here for a single bin,
\begin{equation}
    Q(\vec{p}) = -2\ln\left[\frac{\mathcal {L}(n; \mu(\vec{p}))}{\mathcal {L}(n; n)}\right],
    \label{eq:chi2_trafo}
\end{equation}
where $n$ is the count of samples in the bin, $\mu$ is the model expectation, which is a function of model parameters, $\vec{p}$. In the constant denominator, $\mu$ is replaced by $n$. If the model is correct, $Q$ follows a chi-square distribution with $n_\text{dof}$ degrees of freedom in the asymptotic limit of infinite sample size, where $n_\text{dof}$ is the difference between the number of bins and the number of fitted parameters. In the following, we will sloppily refer to $Q(\vec{p})$ as the \enquote{likelihood}, although it is a monotonic function of a likelihood.

In the likelihoods that are considered here, the nuisance parameters from the templates add a balance of zero to $n_\text{dof}$, since each bin with a simulated count has one corresponding nuisance parameter. Therefore, $n_\text{dof}$ of the total likelihood is still given by the difference of the number of bins and the number of yields.

\section{Derivation of the approximate likelihood}

\eq{chi2_trafo} applied to the exact likelihood described by Barlow and Beeston gives
\begin{equation}
    \begin{aligned}
    Q &= 2 \big(\mu - n - n (\ln \mu - \ln n)\big) \\
    & \quad+ 2 \sum_k \xi_k - a_k - a_k (\ln \xi_k - \ln a_k),
    \end{aligned}
    \label{eq:exact}
\end{equation}
where $\mu = \sum_k y_k \xi_k / M_k$ is the bin expectation, $\xi_k$ is the unknown amplitude of the $k$-th component, $a_k$ is the observed simulated count, $y_k$ is the yield of the $k$-th component, and $M_k$ is a normalization computed by adding all $a_k$ from different bins. In all calculations below, the yields $y_k$ are treated as constants, since the partial problem of finding optimal nuisance parameters for given values of $y_k$ is considered.

Without loss of generality, the template amplitudes can be parameterized as $\xi_k = a_k \beta_k$, since the extremum of the likelihood is also invariant to monotonic transformations of the parameters. The factors $\beta_k$ tend to 1 in the asymptotic limit of an infinite simulated sample. The central approximation of Conway is to set $\beta_k \approx \beta$; the component factors are replaced by a single factor,
\begin{equation}
    \mu = \sum_k \frac{y_k \xi_k}{M_k} = \sum_k \frac{y_k \beta_k a_k}{M_k}
    \approx \beta \underbrace{\sum_k \frac{y_k a_k}{M_k}}_{\mu_0}.
    \label{eq:mu}
\end{equation}
This approximation is valid in the limit where $\mu$ is dominated by a single component. Applied to \eq{exact}, one gets
\begin{equation}   
    \begin{aligned}
    Q & \approx \underbrace{2 (\beta \mu_0 - n - n (\ln (\beta \mu_0) - \ln n))}_{Q_p(n; \beta \mu_0)} \\
    & \quad + 2 \sum_k \beta a_k - a_k - a_k (\ln (\beta a_k) - \ln a_k) 
    \\
      & = Q_p(n; \beta \mu_0) + 2 \sum_k (\beta - 1) a_k - a_k \ln \beta                                                                                            \\
      & = Q_p(n; \beta \mu_0) + 2 a \big((\beta - 1) - \ln \beta\big),
      \end{aligned}
       \label{eq:intermediate}
\end{equation}
with $a = \sum_k a_k$. At this point, the derivation branches. In the next part, the equivalent of Conway's likelihood is derived. In the part after that, the new approximation is derived.

\subsection{Conway's approximation}

The first part $Q_p(n; \beta \mu_0)$ of \eq{intermediate} already corresponds to Conway's likelihood; one only needs to modify the second part. The logarithm $\ln \beta$ is approximated by a Taylor series around $\beta = 1$ to second order. One obtains
\begin{equation}
    \begin{aligned}
    Q & = Q_p(n; \beta \mu_0) + 2 a \big((\beta - 1) - \ln \beta\big)                                 \\
      & \approx Q_p(n; \beta \mu_0) + 2 a \big((\beta - 1) - (\beta - 1) \\
      &\quad + \frac12 (\beta - 1)^2\big) \\
      & = Q_p(n; \beta \mu_0) + a (\beta - 1)^2.
     \end{aligned}
\end{equation}
This approximation is valid in the asymptotic limit $a \to \infty$, but in practice that is at odds with the fact that the simulation sample is frequently smaller than the data sample, therefore, $a$ may be small in some bins with nonzero $n$.

In Conway's likelihood, the second term is divided by the variance $V(\beta)$ of $\beta$, not multiplied by $a$. This is an effective way to counteract the limitation introduced by setting $\beta_k \approx \beta$. To better understand this, it is demonstrated that $V(\beta) \approx 1/a$ when a single component is dominant. Recall the definition $\beta = \mu / \mu_0$, where $\mu_0$ is considered constant. One finds via error propagation,
\begin{equation}
    V(\beta) = \frac{V(\mu)}{\mu_0^2}.
\end{equation}
With
\begin{equation}
    V(\mu) = \sum_k \frac{y_k^2 \, V(\xi_k)}{M_k^2} .
    \label{eq:v_mu}
\end{equation}
With the plug-in estimate $V(\xi_k) = \xi_k \approx a_k$ (the $\xi_k$ are Poisson distributed), one obtains
\begin{equation}
    V(\beta) = \frac{\sum_k \frac{y_k^2}{M_k^2} a_k}{\left(\sum_k \frac{y_k}{M_k} a_k \right)^2}.
    \label{eq:var_beta_conway}
\end{equation}
In the limit where one of the components is dominant, $a_k \approx a$ (the same limit in which the central approximation $\beta_k \approx \beta$ is valid), one gets
\begin{equation}
    V(\beta) \approx \frac{\sum_k \frac{y_k^2}{M_k^2} a}{\left(\sum_k \frac{y_k}{M_k} a \right)^2} = \frac{1}{a}.
\end{equation}
Hence, the equivalent of Conway's likelihood is obtained,
\begin{equation}
    Q = Q_p(n; \beta \mu_0) + \frac{(\beta - 1)^2}{V(\beta)}.
    \label{eq:conway}
\end{equation}
In practice, $V(\beta)$ is computed with \eq{var_beta_conway}, which in the limit of large simulation samples is also correct when more than one component is dominant.

As shown by Conway \cite{Conway:2011in}, an estimate for $\beta$ can be obtained by solving the score function $\partial Q/\partial \beta = 0$ based on \eq{conway}, which leads to a quadratic equation for $\beta$, which has only one valid solution. In summary, Conway's likelihood is correct in the limit of large simulation samples and an approximation to the exact Barlow-Beeston likelihood for small samples.

\subsection{Alternative approximation}

We try to find a better approximation for small simulated samples than Conway's likelihood. Starting again from \eq{intermediate}, the score function $\partial Q/\partial \beta = 0$ is directly solved. One obtains without any further approximation,
\begin{equation}
\begin{aligned}
    \mu_0 - \frac n \beta + a \left(1 - \frac 1 \beta \right) = 0 \; \Rightarrow \; \beta = \frac{n + a}{\mu_0 + a}.
    \end{aligned}
    \label{eq:beta_solution}
\end{equation}
This formula for $\beta$ is even simpler than the one derived by Conway, and can be easily interpreted. If $a \ll n$, $\beta$ adjusts $\mu_0$ to $n$, irrespective of the actual value of $\mu_0$. The bin provides no information on the component yields, which enter only through the potential tension between $\mu_0$ and $n$. For $a \to \infty$, $\beta$ goes to $1$ and $Q \to Q_p(n;\mu_0)$, which is identical to the exact likelihood for the case when the templates are exactly known.

The alternative likelihood in \eq{intermediate} treats data and simulation symmetrically, which becomes more clear after a few transformations,
\begin{equation}
    \begin{aligned}
        Q & = Q_p(n; \beta \mu_0) + 2 a \big((\beta - 1) - \ln \beta \big)                 \\
          & = Q_p(n; \beta \mu_0)  \\
          & \quad + 2\big((\beta a - a) - a (\ln \beta + \ln a - \ln a)\big)  \\
          & = Q_p(n; \beta \mu_0) + 2\big((\beta a - a) - a (\ln (\beta a) - \ln a)\big)      \\
          & = Q_p(n; \beta \mu_0) + Q_p(a; \beta a).
        \label{eq:alternative}
    \end{aligned}
\end{equation}

To summarize, this alternative approximate likelihood is derived from the exact likelihood without using a Gaussian approximation, in contrast to Conway's likelihood. It describes both the simulated and the data sample with Poisson statistics. As shown by Barlow and Beeston \cite{Barlow:1993dm} and elsewhere, maximum-likelihood estimation based on Poisson statistics generally performs better than a likelihood based on a Gaussian approximation for Poisson-distributed data. The alternative likelihood should therefore perform better in fits where the simulated sample used to produce the templates is small.

Bins in which $a$ is zero should be ignored in the calculation, since they do not contribute anything to the yields. In practice, one can replace $a$ with a tiny number, so that \eq{beta_solution} and \eq{alternative} always produces a valid number instead of dividing by zero or taking the logarithm of zero.

\section{Weighted templates and weighted data}

In practice, the data and the simulation samples may be weighted. Simulations are often weighted to reduce discrepancies between the simulated and the real experiment, or as a form of importance sampling. Data may be weighted to correct losses from finite detection efficiency. In both cases, a count is replaced by a sum of weights, $\sum_i w_i$. Barlow and Beeston discuss this case in their paper and show how the exact likelihood can be adapted, but their solution is not complete since it does not handle the additional variance introduced by having weights of varying size. A better approximate solution is given here, which is correct in the asymptotic limit of large samples.

\eq{conway} and \eq{alternative} are adapted by replacing $Q_P$ with the approximate likelihood for sums of independent weights, described by Bohm and Zech \cite{Bohm:2013gla}. In their approach, the sum of weights $\sum_i w_i$ and the prediction $\mu$ are scaled with a factor $s = \sum_i w_i / \sum_i w_i^2$. With the so-called effective count, $n_\text{eff} = (\sum_i w_i)^2 / \sum_i w_i^2$, the full replacement becomes
\begin{equation}
    Q_P(n; \mu) \rightarrow Q_P\left(n_\text{eff}; s \mu \right).
    \label{eq:bz}
\end{equation}
This replacement can be applied to both parts of \eq{alternative},
\begin{equation}
    Q = Q_P(n_\text{eff}; \beta s \mu_0) + Q_P(a_\text{eff}; \beta a_\text{eff}),
    \label{eq:weighted_alternative}
\end{equation}
where $a_\text{eff}$ is the equivalent of $n_\text{eff}$ computed from the weights in the simulation.

These replacements also modify the solution for $\beta$. By identifying the corresponding variables in \eq{alternative} and \eq{weighted_alternative}, one finds
\begin{equation}
    \beta = \frac{n_\text{eff} + a_\text{eff}}{s \mu_0 + a_\text{eff}}.
\end{equation}

In the case of Conway's likelihood, weighted data is handled in the same way. Weighted templates only affect the calculation of $V(\beta)$, which is again derived from \eq{v_mu}, but now with variance $V(\xi_k) = \sum_i w_i^2$, where $w_i$ are the weights of template $k$.

\section{Improved alternative approximation}

The alternative approximate likelihood derived so far is valid in the limit that each bin is dominated by a single template component. If there is only one template component, the approximation is exact. This insight combined with formulas from the previous section leads to an improved approximation that extends the validity, inspired by Conway's approach to compute $V(\beta)$ with \eq{var_beta_conway}.

The expected count $\mu$ per bin, given by \eq{mu}, can be regarded as a single template component in the previous sense. It is a random variable, because it is computed from Poisson-distributed template values. Random samples of $\mu$ are not Poisson-distributed, their variance is given by \eq{v_mu}. Following again Bohm and Zech \cite{Bohm:2013gla}, an approximate likelihood for $\mu$ is
\begin{equation}
    \begin{aligned}
    &Q_P (s_\mu \mu_0; s_\mu \mu) = Q_P(s_\mu \mu_0; s_\mu \beta \mu_0) \\
    & = Q_P(\mu_\text{eff}; \beta \mu_\text{eff}),
    \end{aligned}
\end{equation}
with $\mu_\text{eff} = \mu_0^2 / V_\mu$.
The approximate likelihood becomes
\begin{equation}
    Q = Q_P(n_\text{eff}; \beta s \mu_0) + Q_P(\mu_\text{eff}; \beta \mu_\text{eff}),
    \label{eq:alternative2}
\end{equation}
and the solution for $\beta$ is
\begin{equation}
    \beta = \frac{n_\text{eff} + \mu_\text{eff}}{s \mu_0 + \mu_\text{eff}}.
\end{equation}
In the limit that a single template component dominates $\mu_0$, \eq{alternative2} reduces to \eq{alternative}, but it also produces very good results if more than one component is dominant.

\section{Toy study}

The properties of the estimated yields, obtained with the likelihoods presented here, are studied in two toy examples. The yields are estimated by minimizing the likelihoods given by \eq{exact} (Barlow-Beeston), by \eq{conway} (Conway), and \eq{alternative2} (this work). Of interest is the bias of the estimates and the bias of the estimated uncertainty of the yields. Both biases should be small.

The minimization is performed with the MINUIT \cite{James:1975dr} algorithm MIGRAD as implemented in \texttt{iminuit} \cite{iminuit}. In the case of \eq{exact}, the nuisance parameters are found by the Barlow-Beeston algorithm from the reference implementation \texttt{TFractionFitter} in the ROOT framework \cite{rene_brun_2019_3895860}.

\begin{figure}[tbp]
    \includegraphics[width=\linewidth]{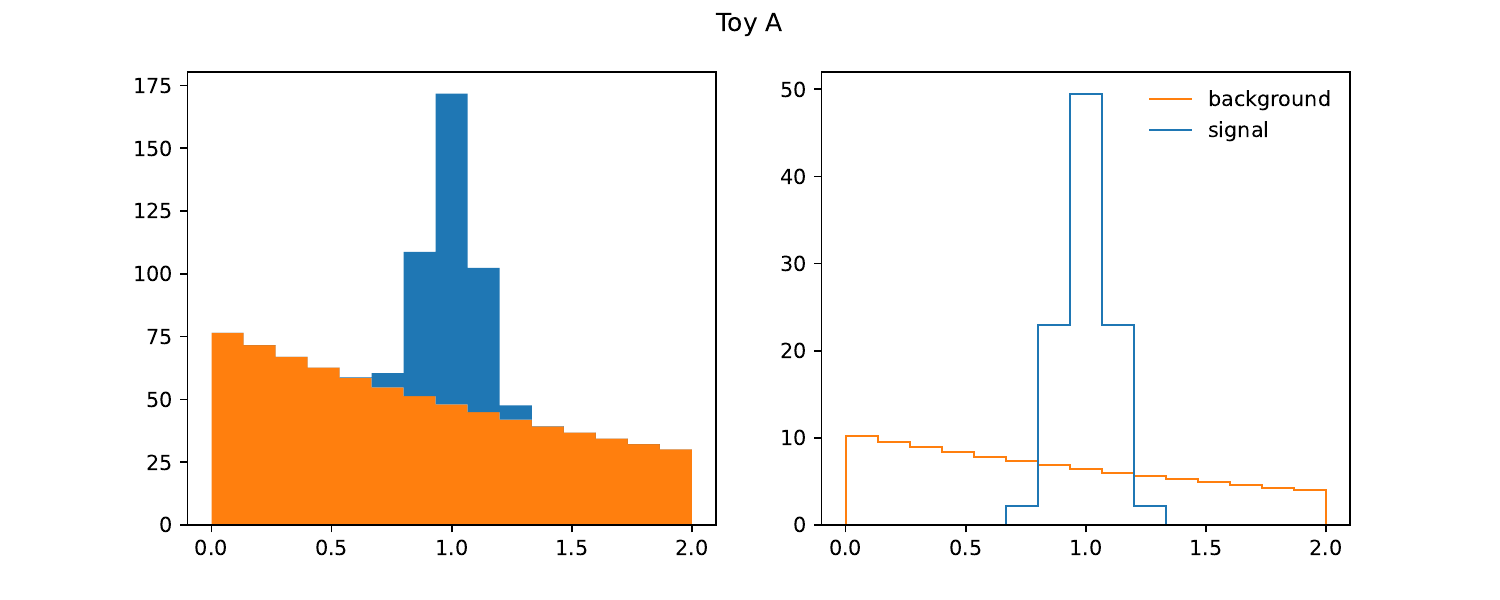}
    \includegraphics[width=\linewidth]{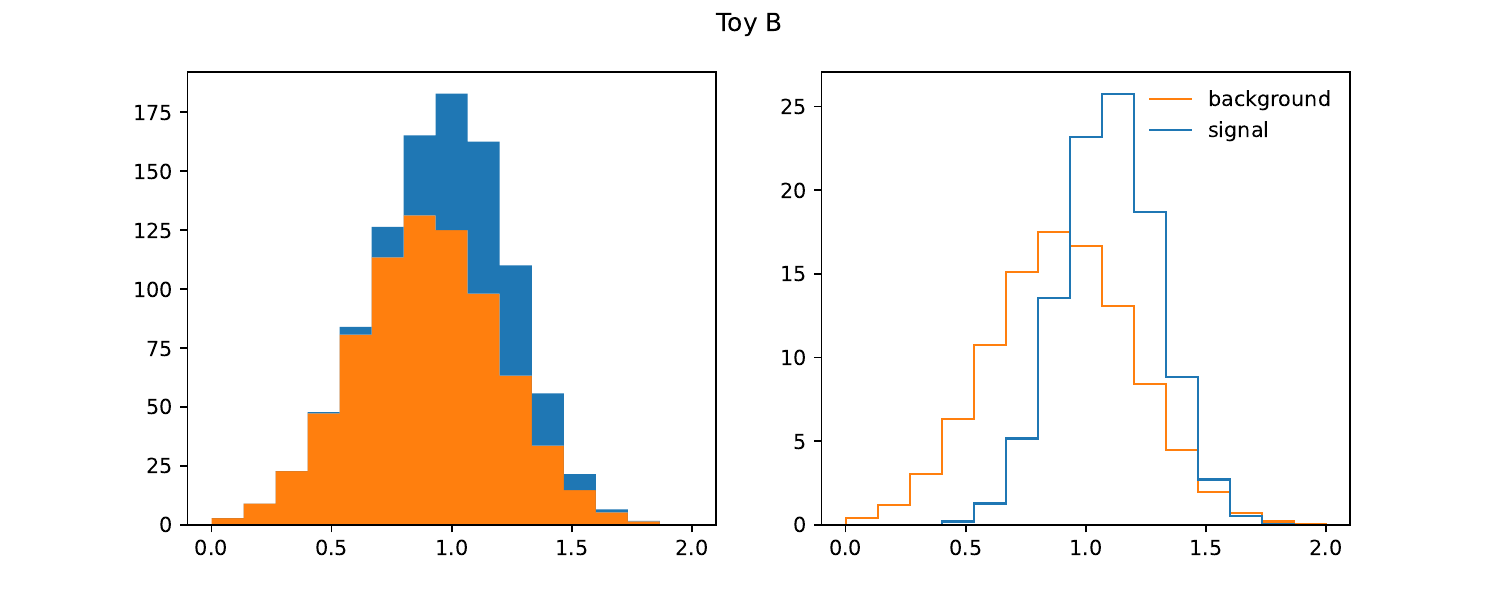}
    \caption{The data distribution is shown on the left-hand side, the corresponding templates on the right-hand side. Shown in the top (bottom) row is example A (B). The expected values per bin are shown instead of a random sample. The integral of the data distribution is 1000. The integral of each template, $N_\text{mc}$, is varied ($N_\text{mc} = 100$ in this case).}\label{fg:example}
\end{figure}

\begin{table*}[tbp]
    \centering
    \caption{Parameters of the distributions used in the toy simulation. The signal and background distributions are truncated to the interval $[0, 2]$.}
    \begin{tabular}{l l l}
                & Signal                                                & Background                          \\
        \hline
        Case A  & normal, $\mu = 1$, $\sigma = 0.1$                     & exponential, $\mu = 2$              \\
        Case B  & normal, $\mu = 1.1$, $\sigma = 0.2$                   & normal, $\mu = 0.9$, $\sigma = 0.3$ \\
        Weights & \multicolumn{2}{c}{uniform in the interval $[0, 10]$}
    \end{tabular}
    \label{tb:toy}
\end{table*}

Both toy examples consist of two mixed components, with the parameters listed in \tb{toy}. In case A, a normally distributed signal is mixed with a comparably flat exponentially distributed background. In this example, most bins are dominated by a single component. Two normally distributed components are mixed for case B so that most bins are not clearly dominated by one component. Case B is a more difficult problem than case A. The expected signal yield is 250, whilst the expected background yield is 750. The expected yield per component template is set to $N_\text{mc}$, where $N_\text{mc}$ varies between sets of experiments. When a new toy experiment is simulated, all yields are randomly drawn from the Poisson distribution. Generated samples are sorted into histograms with 15 equidistant bins over the interval $[0, 2]$. The expected counts per bin are shown in \fg{example} for both cases. A template fit is then performed based on these inputs. As a variation, this simulation is repeated with weighted templates. To simulate weighted templates, $k$ random weights are drawn for each bin in a template, where $k$ is the Poisson-distributed count in that bin generated by the algorithm described above. The random weights are drawn from a uniform distribution in the interval $[0, 10]$. To clearly illustrate the difference in performance between the Barlow-Beeston likelihood and the new likelihoods described here, the weight distribution is chosen so that the sum of weights, which replaces the bin count, has a variance larger than a Poisson distribution by a factor 6 to 7.

Independent samples are generated and fitted 2000 times for each value of $N_\text{mc}$ to measure the bias of the estimate itself and the bias of its variance estimate. Sets of experiments are run for $N_\text{mc} = $ 100, 1000, 10000. It is expected that the results converge as the sample used to compute the templates grows. To judge the performance, the pull distribution of the estimated signal yield is computed, where the pull is defined as
\begin{equation}
    z = (\hat s - s)/ \hat V_s^{1/2},
\end{equation}
with true signal yield $s$, estimate $\hat s$, and estimated variance $\hat V_s$ (obtained by MINUIT's HESSE algorithm) for $\hat s$. The performance is indicated by the degree of agreement of the mean of $z$ with 0 and the standard deviation of $z$ with 1.

\begin{figure*}
    \includegraphics[width=\linewidth]{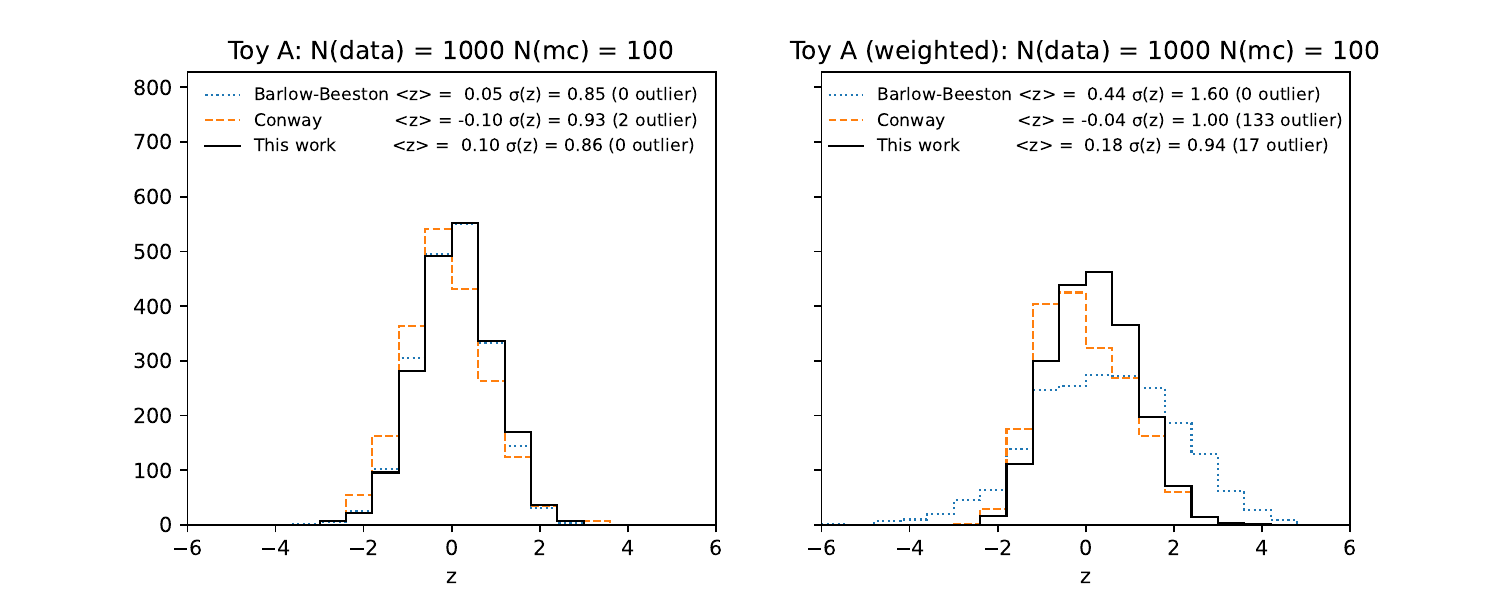}
    \includegraphics[width=\linewidth]{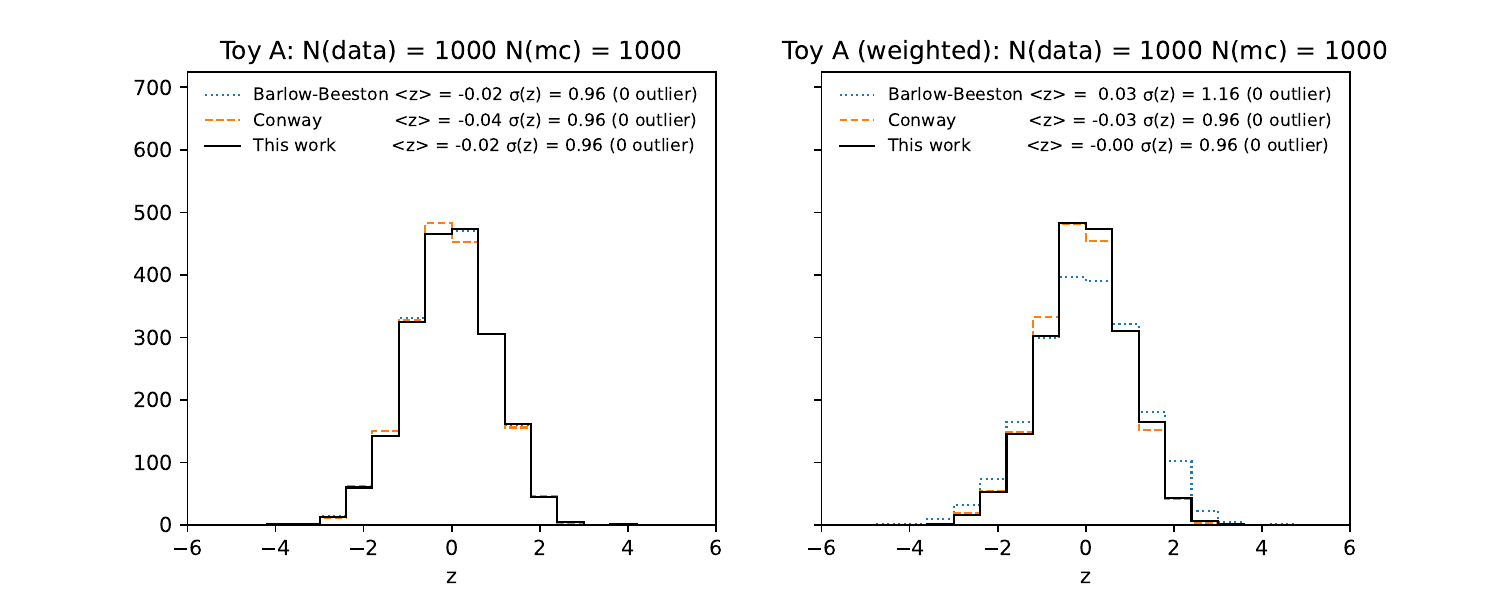}
    \includegraphics[width=\linewidth]{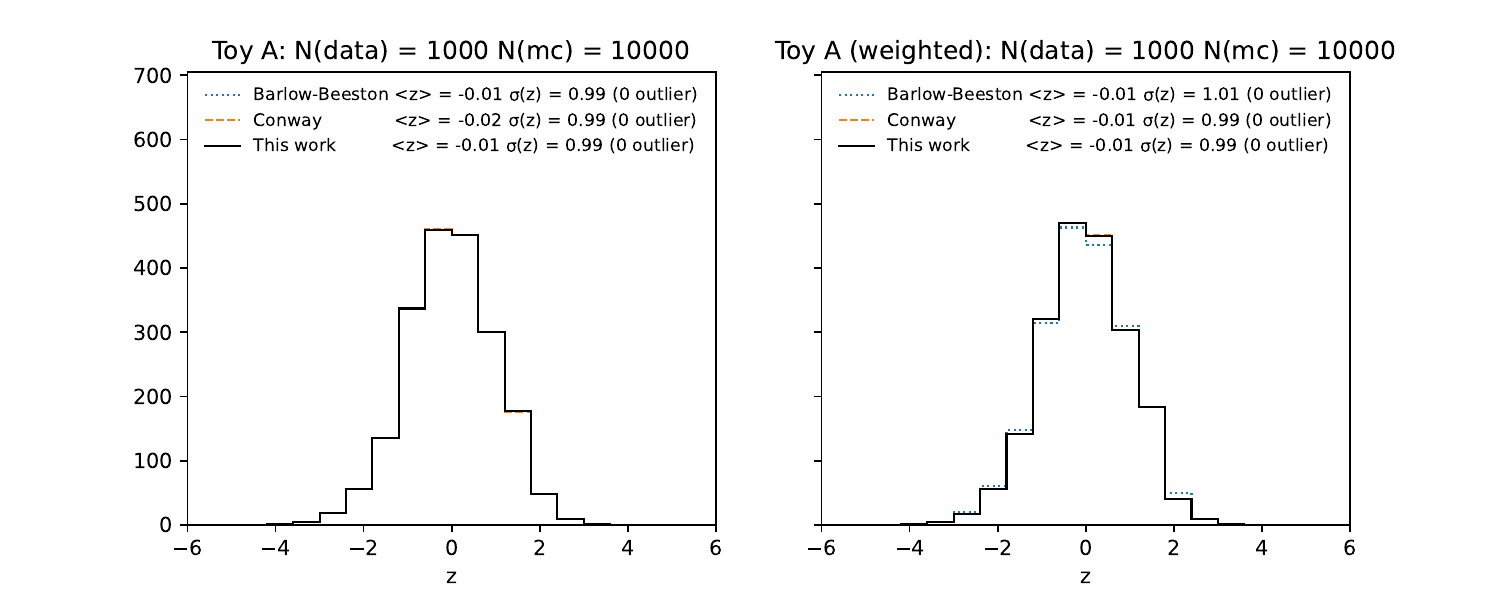}
    \caption{Pull distributions for toy example A for the estimates obtained from maximizing the exact likelihood (Barlow-Beeston), the approximate likelihoods of Conway, and the one derived in this work. Shown on the left-hand side are fits with Poisson-distributed templates, shown on the right-hand side are fits with weighted templates as described in the text. Outliers with $|z| > 5$ are not included in the calculation of mean and standard deviation in the legend. The number of outliers is included in the legend.}
    \label{fg:study_a}
\end{figure*}

\begin{figure*}
    \includegraphics[width=\linewidth]{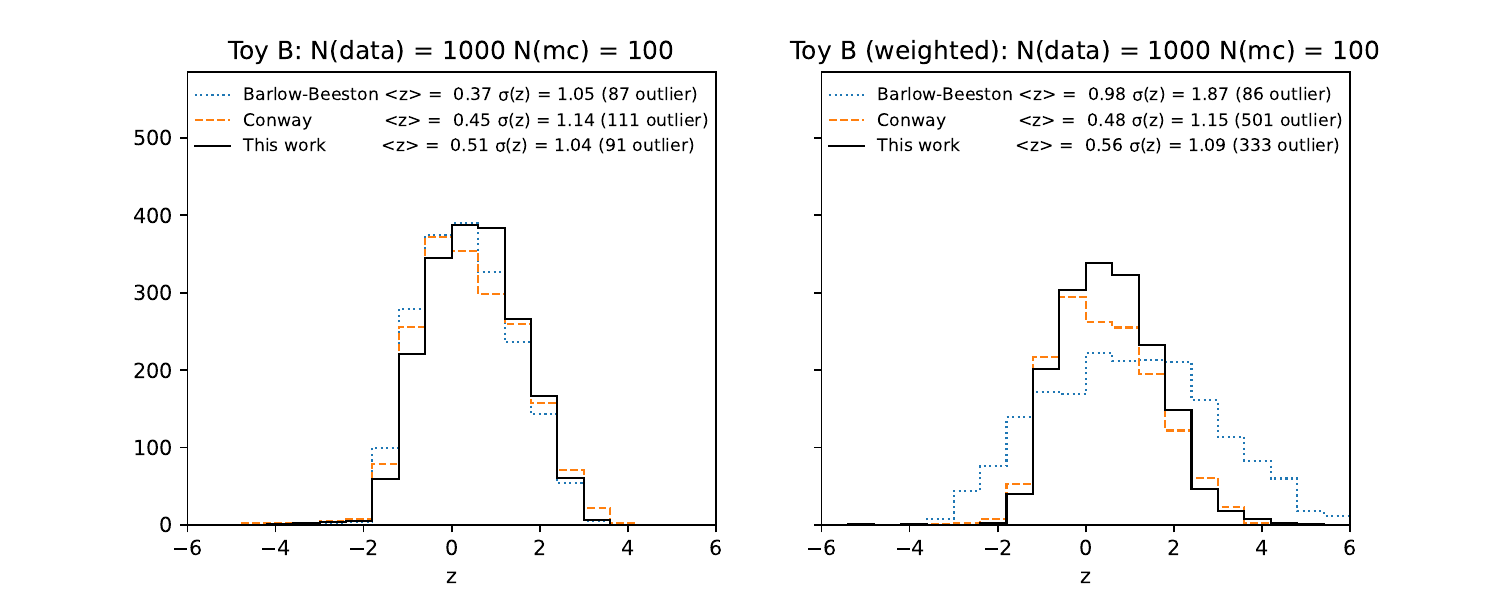}
    \includegraphics[width=\linewidth]{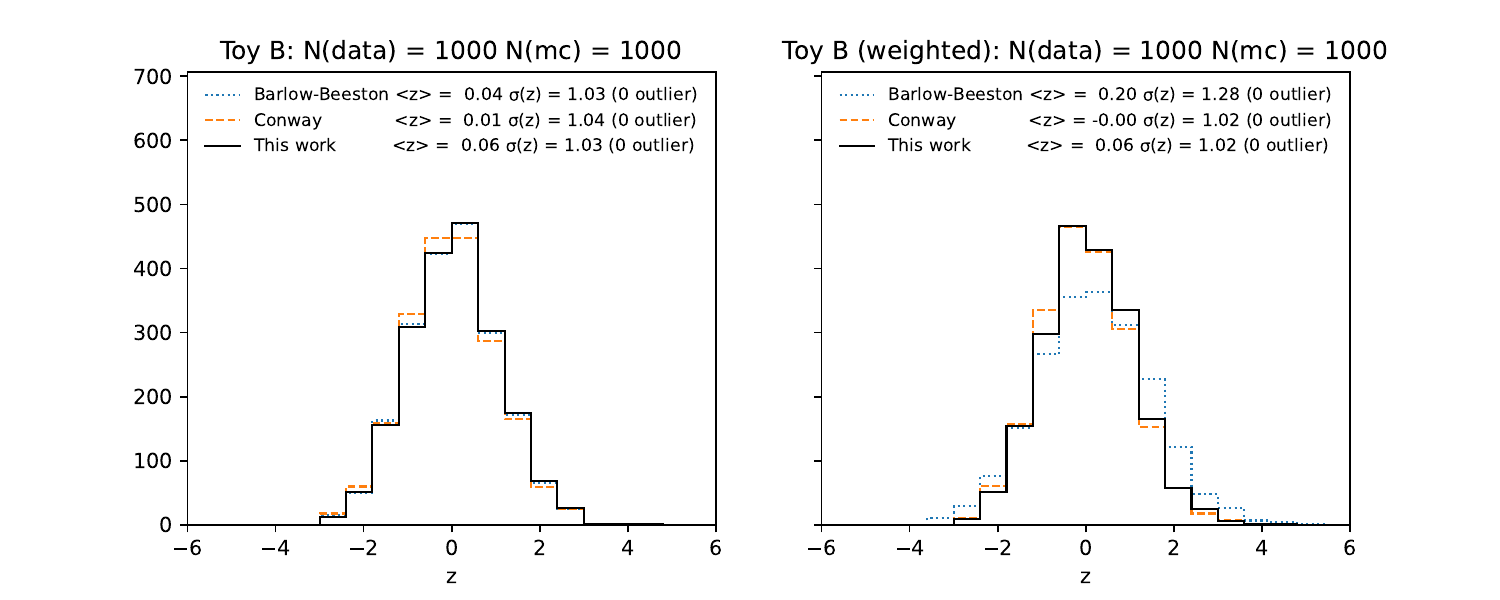}
    \includegraphics[width=\linewidth]{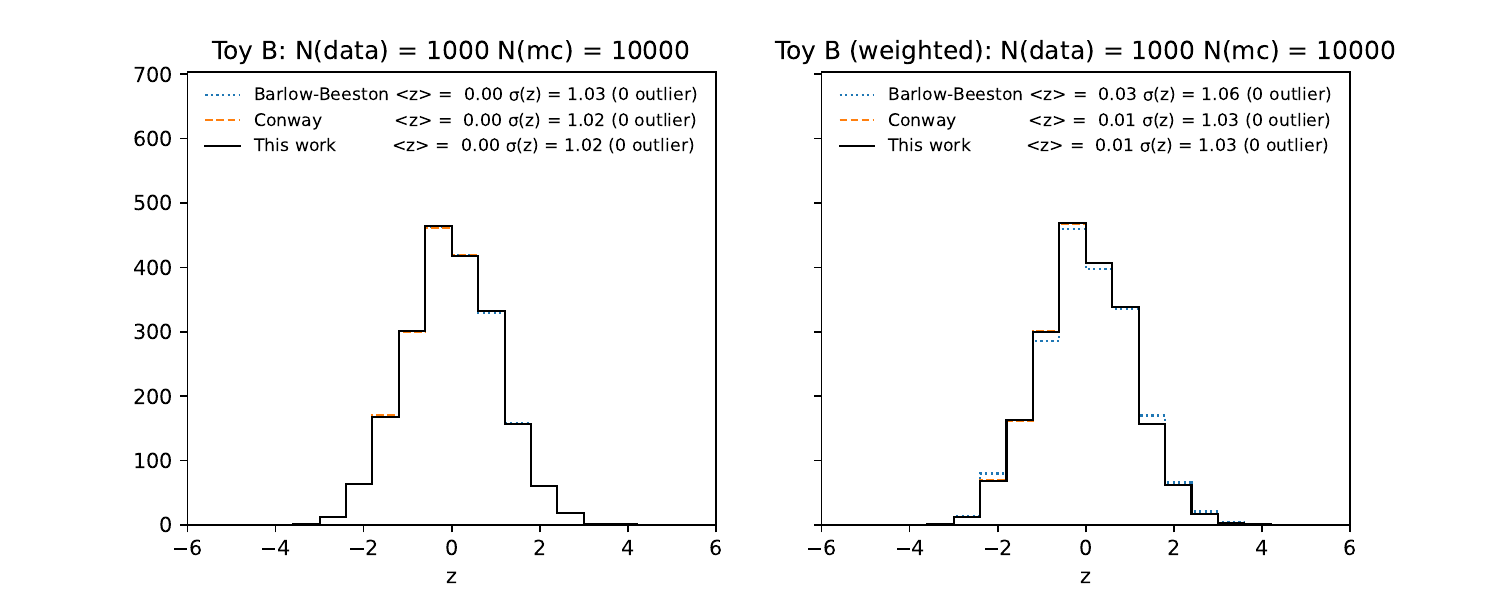}
    \caption{Plots for toy example B analog to the ones in \fg{study_a}}
    \label{fg:study_b}
\end{figure*}

The results are shown in \fg{study_a} and \ref{fg:study_b}. For large simulation samples, all likelihoods produce compatible results, since the uncertainty contributed by the simulated templates becomes negligible.
The performance of all likelihoods on Poisson-distributed templates is also comparable if outliers with $|z| > 5$ are removed. In the strongly mixed case B, many outliers are produced for $N_\text{mc} = 100$, and signal amplitudes are fairly biased. This corresponds to the intuition that it is difficult to estimate a small signal component over a large background unless the shapes of the signal and background are well separated. It is important to minimize the number of outliers because they cannot be removed in practice. The Barlow-Beeston likelihood produces the smallest number of outliers, closely followed by the new approximate likelihood. Conway's likelihood produces the largest number of outliers. In fits using weighted templates, the Barlow-Beeston likelihood underestimates the true variance of the fitted yields, while the two generalized approximate likelihoods match the true variance.

\begin{figure}
    \centering
    \includegraphics[width=\linewidth]{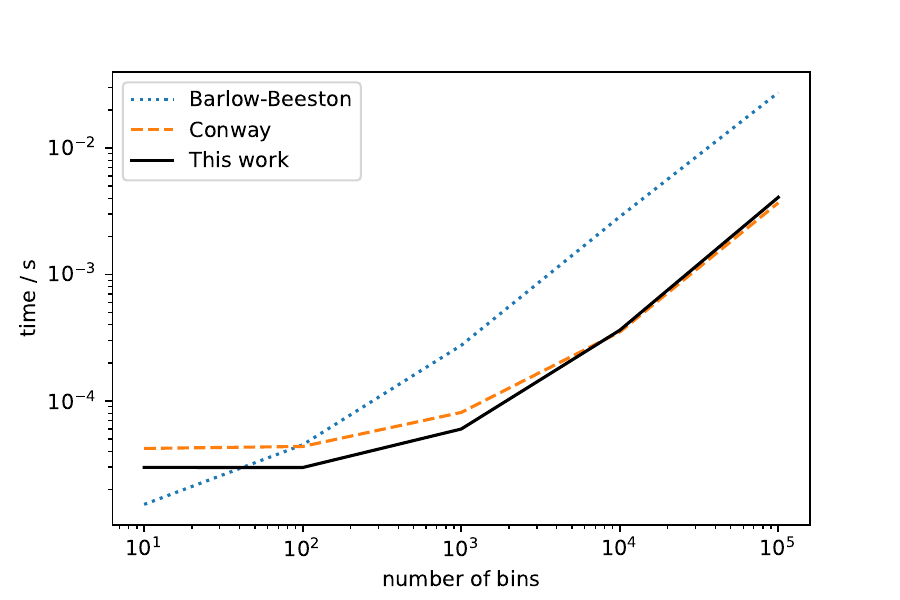}
    \caption{Time required to evaluate the likelihoods described in the text; smaller is better. The likelihoods were evaluated on the toy example A. The number of bins in the data and template histograms was varied while keeping the bin density constant.}
    \label{fg:timing}
\end{figure}

The runtime of a fit is dominated by the time required to evaluate the likelihood and therefore it is interesting to compare evaluation times. These times are specific to the problem, computing platform, and the implementation. The Barlow-Beeston algorithm is implemented in C++, while the approximate likelihoods discussed in this work are implemented in Python. Code execution in Python is orders of magnitude slower than in C++, but the use of the Numpy library \cite{harris2020array} makes the performance of the Python implementations competitive with C++. The approximate likelihoods outperform the Barlow-Beeston algorithm as the number of bins is larger than about 200, because the internal computation of the nuisance parameters is much less time-consuming. In practice, a bin number of 200 is easily exceeded when the distributions are multidimensional. The evaluation time of the two approximate likelihoods becomes comparable as the number of bins grows, while for small number of bins, the new approximate likelihood derived in this work is 25\,\% faster than Conway's likelihood.

\section{Conclusions}

Conway's approximate likelihood and a new alternative approximate likelihood were derived from the exact likelihood described by Barlow and Beeston. The new approximate likelihood treats data and simulation symmetrically as Poisson distributed, while Conway used a different treatment for the simulation. The new approximate likelihood is expected to perform better when the simulation sample is small. The two approximate likelihoods were generalized to approximately describe weighted data and/or weighted templates, and to correctly take the increased fluctuations into account that weighted histograms have when the weights vary in size. This goes beyond the treatment that Barlow and Beeston describe in their 1993 paper for weighted templates, where the increased fluctuations are not handled. All likelihood functions were transformed so that the minimum value is asymptotically chi-square distributed following the approach described by Baker and Cousins, which turns the minimum value into a goodness-of-fit test statistic.

A study of two toy examples showed that yields obtained with fits with Poisson-distributed templates, the Barlow-Beeston likelihood, Conway's likelihood, and the new approximate likelihood derived in this work have comparable performance if outliers are removed. However, outliers cannot be removed in a real analysis therefore it is important to minimize outliers. The new approximate likelihood derived here produced slightly more outliers than the Barlow-Beeston likelihood, but fewer outliers than Conway's likelihood. In fits with weighted templates, the Barlow-Beeston likelihood underestimated the true variance of the yields. Only fits with the generalized approximate likelihoods derived in this work estimate the variance well.

The approximate likelihoods derived in this work have a significantly smaller computational cost than the Barlow-Beeston likelihood when the number of bins in the distributions is large. This benefits the runtime of a fit, which is usually dominated by the evaluation time of the likelihood.

The generalized approximate likelihoods developed in this paper are available in the \texttt{iminuit} library \cite{iminuit}, and can be used by creating instances of the class \texttt{iminuit.cost.BarlowBeestonLite}.

\section*{Declarations}
\subsection*{Funding}
HD acknowledges funding by the Deutsche Forschungsgemeinschaft (DFG, German Research Foundation) -- project no. 449728698.

\subsection*{Competing interests}
The authors declare that they have no known competing financial interests or personal relationships that could have appeared to influence the work reported in this paper.

\subsection*{Code availability}
 The algorithm code is available on the iminuit website: \url{https://iminuit.readthedocs.io/en/stable/reference.html#iminuit.cost.BarlowBeestonLite}

\subsection*{Data availability}
The datasets generated during and/or analysed during the current study are available from the corresponding author on reasonable request.

\bibliography{main}

\end{document}